\documentclass{PoS}

\title{Probing Efficient Cosmic-Ray Acceleration in Young Supernovae }

\ShortTitle{Probing Efficient Cosmic-Ray Acceleration in Young Supernovae }

\author{\speaker{V.~V.~Dwarkadas}\\
  Department of Astronomy and Astrophysics, University of Chicago, 5640 S Ellis Ave, Chicago, IL 60637\\
        E-mail: \email{vikram@oddjob.uchicago.edu}}

\author{M. Renaud,  A. Marcowith\\
        Laboratoire Univers et particules de Montpellier, Universit\'e Montpellier II/CNRS, place E. Bataillon, cc072, 34095 Montpellier, France}

\author{V. Tatischeff\\
        Centre de Sciences Nucl\'eaires et de Sciences de la Mati\`ere, IN2P3/CNRS and Univ Paris-Sud, 91405 Orsay, France\\}

\abstract{The formation of a core collapse supernovae (SNe) results in
  a fast (but non- or mildly-relativistic) shock wave expanding
  outwards into the surrounding medium. The medium itself is likely
  modified due to the stellar mass-loss from the massive star
  progenitor, which may be Wolf-Rayet stars (for Type Ib/c SNe), red
  supergiant stars (for type IIP and perhaps IIb and IIL SNe), or some
  other stellar type. The wind mass-loss parameters determine the
  density structure of the surrounding medium. Combined with the
  velocity of the SN shock wave, this regulates the shock acceleration
  process. In this article we discuss the essential parameters that
  control the particle acceleration and gamma-ray emission in SNe,
  with particular reference to the Type IIb SN 1993J. The shock wave
  expanding into the high density medium leads to fast particle
  acceleration, giving rise to rapidly-growing plasma instabilities
  driven by the acceleration process itself. The instabilities grow
  over intraday timescales. This growth, combined with the interplay
  of non-linear processes, results in the amplification of the
  magnetic field at the shock front, which can adequately account for
  the magnetic field strengths deduced from radio monitoring of the
  source. The maximum particle energy can reach, and perhaps exceed, 1
  PeV, depending on the dominant instability.  The gamma-ray signal is
  found to be heavily absorbed by pair production process during the
  first week after the outburst. We derive the time dependent particle
  spectra and associated hadronic signatures of secondary particles
  (gamma-ray, leptons and neutrinos) arising from proton proton
  interactions. We find that the Cherenkov Telescope Array (CTA)
  should be able to detect objects like SN 1993J above 1 TeV.  We
  predict a low neutrino flux above 10 TeV, implying a detectability
  horizon with current or planned neutrino telescopes of 1 Mpc.}

\FullConference{The 34th International Cosmic Ray Conference \\
		 30 July- 6 August, 2015\\
		 The Hague, The Netherlands}

\begin{document}
\newcommand{\vper}{\mbox{${v_{\perp}}$}}
\newcommand{\vpar}{\mbox{${v_{\parallel}}$}}
\newcommand{\uper}{\mbox{${u_{\perp}}$}}
\newcommand{\vperout}{\mbox{${{v_{\perp}}_{o}}$}}
\newcommand{\uperout}{\mbox{${{u_{\perp}}_{o}}$}}
\newcommand{\vperin}{\mbox{${{v_{\perp}}_{i}}$}}
\newcommand{\uperin}{\mbox{${{u_{\perp}}_{i}}$}}
\newcommand{\upar}{\mbox{${u_{\parallel}}$}}
\newcommand{\uparout}{\mbox{${{u_{\parallel}}_{o}}$}}
\newcommand{\vparout}{\mbox{${{v_{\parallel}}_{o}}$}}
\newcommand{\uparin}{\mbox{${{u_{\parallel}}_{i}}$}}
\newcommand{\vparin}{\mbox{${{v_{\parallel}}_{i}}$}}
\newcommand{\dout}{\mbox{${\rho}_{o}$}}
\newcommand{\din}{\mbox{${\rho}_{i}$}}
\newcommand{\da}{\mbox{${\rho}_{1}$}}
\newcommand{\mfast}{\mbox{$\dot{M}_{f}$}}
\newcommand{\mslow}{\mbox{$\dot{M}_{a}$}}
\newcommand{\beqn}{\begin{eqnarray}}
\newcommand{\eeqn}{\end{eqnarray}}
\newcommand{\be}{\begin{equation}}
\newcommand{\ee}{\end{equation}}
\newcommand{\noi}{\noindent}
\newcommand{\ftheta}{\mbox{$f(\theta)$}}
\newcommand{\gtheta}{\mbox{$g(\theta)$}}
\newcommand{\ltheta}{\mbox{$L(\theta)$}}
\newcommand{\stheta}{\mbox{$S(\theta)$}}
\newcommand{\utheta}{\mbox{$U(\theta)$}}
\newcommand{\xitheta}{\mbox{$\xi(\theta)$}}
\newcommand{\vs}{\mbox{${v_{s}}$}}
\newcommand{\ro}{\mbox{${R_{0}}$}}
\newcommand{\pa}{\mbox{${P_{1}}$}}
\newcommand{\va}{\mbox{${v_{a}}$}}
\newcommand{\vo}{\mbox{${v_{o}}$}}
\newcommand{\vp}{\mbox{${v_{p}}$}}
\newcommand{\vw}{\mbox{${v_{w}}$}}
\newcommand{\vf}{\mbox{${v_{f}}$}}
\newcommand{\lprime}{\mbox{${L^{\prime}}$}}
\newcommand{\uprime}{\mbox{${U^{\prime}}$}}
\newcommand{\sprime}{\mbox{${S^{\prime}}$}}
\newcommand{\xiprime}{\mbox{${{\xi}^{\prime}}$}}
\newcommand{\mdot}{\mbox{$\dot{M}$}}
\newcommand{\msun}{\mbox{$M_{\odot}$}}
\newcommand{\yr}{\mbox{${\rm yr}^{-1}$}}
\newcommand{\kms}{\mbox{${\rm km} \;{\rm s}^{-1}$}}
\newcommand{\lambdav}{\mbox{${\lambda}_{v}$}}
\newcommand{\lequ}{\mbox{${L_{eq}}$}}
\newcommand{\eqpratio}{\mbox{${R_{eq}/R_{p}}$}}
\newcommand{\ra}{\mbox{${r_{o}}$}}
\newcommand{\bfig}{\begin{figure}[h]}
\newcommand{\efig}{\end{figure}}
\newcommand{\tone}{\mbox{${t_{1}}$}}
\newcommand{\done}{\mbox{${{\rho}_{1}}$}}
\newcommand{\dsn}{\mbox{${\rho}_{SN}$}}
\newcommand{\dzero}{\mbox{${\rho}_{0}$}}
\newcommand{\ve}{\mbox{${v}_{e}$}}
\newcommand{\vej}{\mbox{${v}_{ej}$}}
\newcommand{\Mch}{\mbox{${M}_{ch}$}}
\newcommand{\mej}{\mbox{${M}_{e}$}}
\newcommand{\Mst}{\mbox{${M}_{ST}$}}
\newcommand{\dam}{\mbox{${\rho}_{am}$}}
\newcommand{\Rst}{\mbox{${R}_{ST}$}}
\newcommand{\Vst}{\mbox{${V}_{ST}$}}
\newcommand{\Tst}{\mbox{${T}_{ST}$}}
\newcommand{\no}{\mbox{${n}_{0}$}}
\newcommand{\Efif}{\mbox{${E}_{51}$}}
\newcommand{\rsh}{\mbox{${R}_{sh}$}}
\newcommand{\msh}{\mbox{${M}_{sh}$}}
\newcommand{\vsh}{\mbox{${V}_{sh}$}}
\newcommand{\vrev}{\mbox{${v}_{rev}$}}
\newcommand{\rpr}{\mbox{${R}^{\prime}$}}
\newcommand{\mpr}{\mbox{${M}^{\prime}$}}
\newcommand{\vpr}{\mbox{${V}^{\prime}$}}
\newcommand{\tpr}{\mbox{${t}^{\prime}$}}
\newcommand{\cone}{\mbox{${c}_{1}$}}
\newcommand{\ctwo}{\mbox{${c}_{2}$}}
\newcommand{\cthree}{\mbox{${c}_{3}$}}
\newcommand{\cfour}{\mbox{${c}_{4}$}}
\newcommand{\Te}{\mbox{${T}_{e}$}}
\newcommand{\Ti}{\mbox{${T}_{i}$}}
\newcommand{\Ha}{\mbox{${H}_{\alpha}$}}
\newcommand{\Rprime}{\mbox{${R}^{\prime}$}}
\newcommand{\Vprime}{\mbox{${V}^{\prime}$}}
\newcommand{\Tprime}{\mbox{${T}^{\prime}$}}
\newcommand{\Mprime}{\mbox{${M}^{\prime}$}}
\newcommand{\rprime}{\mbox{${r}^{\prime}$}}
\newcommand{\rfprime}{\mbox{${r}_f^{\prime}$}}
\newcommand{\vprime}{\mbox{${v}^{\prime}$}}
\newcommand{\tprime}{\mbox{${t}^{\prime}$}}
\newcommand{\mprime}{\mbox{${m}^{\prime}$}}
\newcommand{\Me}{\mbox{${M}_{e}$}}
\newcommand{\nh}{\mbox{${n}_{H}$}}
\newcommand{\rr}{\mbox{${R}_{2}$}}
\newcommand{\rf}{\mbox{${R}_{1}$}}
\newcommand{\vtwo}{\mbox{${V}_{2}$}}
\newcommand{\vout}{\mbox{${V}_{1}$}}
\newcommand{\dshell}{\mbox{${{\rho}_{sh}}$}}
\newcommand{\dwind}{\mbox{${{\rho}_{w}}$}}
\newcommand{\dslow}{\mbox{${{\rho}_{s}}$}}
\newcommand{\dfast}{\mbox{${{\rho}_{f}}$}}
\newcommand{\vfast}{\mbox{${v}_{f}$}}
\newcommand{\vslow}{\mbox{${v}_{s}$}}
\newcommand{\cc}{\mbox{${\rm cm}^{-3}$}}
\newcommand{\apj}{\mbox{ApJ}}
\newcommand{\apjl}{\mbox{ApJL}}
\newcommand{\apjs}{\mbox{ApJS}}
\newcommand{\aj}{\mbox{AJ}}
\newcommand{\araa}{\mbox{ARAA}}
\newcommand{\nat}{\mbox{Nature}}
\newcommand{\aap}{\mbox{AA}}
\newcommand{\gca}{\mbox{GeCoA}}
\newcommand{\pasp}{\mbox{PASP}}
\newcommand{\mnras}{\mbox{MNRAS}}
\newcommand{\apss}{\mbox{ApSS}}
\newcommand{\ssr}{\mbox{SSR}}
\newcommand{\aapr}{\mbox{AARs}}

\section{Introduction} 
Supernova Remnants (SNRs) are widely accepted as the likely source of
Galactic cosmic rays (CR) up to energies of 3 PeV (the CR ``knee''
\cite{blasi13}).  Per the Hillas confinement criterion \cite{hillas84}
that the Larmor radius of the particle match the source size, only
certain Galactic sources should be able to produce such energetic
particles. These include extended sources with standard interstellar
medium (ISM) magnetic field (MF) values such as massive star clusters
and their superbubbles \cite{bykov01, parizot04, ferrand10}; or
compact sources with more intense magnetic fields, such as young
SNRs. In this work we explore particle acceleration and gamma-ray
radiation in SNe arising from massive star progenitors, with
particular focus on the well-monitored SN 1993J. We consider the
acceleration of particles starting from the days following SN outburst
when the radio luminosity is close to its maximum. We review the main
results of radio observations and modeling in \S \ref{S:Acc}, examine
models of magnetic field amplification (MFA) in SN 1993J in \S
\ref{Su:MFA}, and the implied cosmic ray energies in \S
\ref{S:Max}. Modeling of different particle distribution species in
radio SNe is outlined in \S \ref{S:CRs}. Our calculation of the
gamma-ray radiation is shown in \S \ref{S:GRa}, and the neutrino
signal in \S \ref{S:NEu}.  Some perspectives for other types of SNe,
along with conclusions, are discussed in \S \ref{S:Dis}.

\section{A case study: SN 1993J}
\label{S:Acc}
Our model for SN 1993J is based on work by \cite{vt09}(hereafter T09)
that discusses particle acceleration in SN 1993J. This Type IIb SN
resulted from the explosion of a massive K-supergiant star with an
initial mass $\sim$ 15 $M_{\odot}$ \cite{ms09}, having a B-star binary
companion \cite{foxetal14}. The progenitor star had a mass loss rate
$\sim 3.8 \times 10^{-5} M_{\odot}$ yr$^{-1}$ and a wind velocity
$u_{w} \sim 10$ km s$^{-1}$ (T09). Assuming a constant mass loss rate,
the circumstellar medium (CSM) density scales as $n_{circ}= {\dot{M}
  (1+2X) \over 4\pi r^2 u_w m_H (1+4X)}$ where $X=0.1$ is the Helium
fraction and $m_H$ the hydrogen atom mass. The effective density
downstream of the forward shock, one day after outburst, is $n_{eff}
\simeq 4\times 10^9 \ \rm{cm^{-3}}$ with a shock compression ratio of
4. The shock radius after 1 day has been estimated as $r \simeq
3.5\times 10^{14} \ \rm{cm}$ (T09) from the radio expansion. The shock
propagates into a fully ionized medium \cite{flc96}.

The time dependence of the average MF from the synchrotron emitting
shell is (T09):
\begin{equation}
\label{Eq:MFo}
\langle B \rangle \simeq [2.4 \pm 1 \ \rm{G} ] \times \left( {t \over 100~\rm{days}} \right)^{-1.16 \pm 0.20} \ ,
\end{equation}
giving a MF of order $500$ G after 1 day. This field is consistent
with observations \cite{chandraetal04}.  It is an averaged value over
the synchrotron shell and does not necessarily represent the MF
produced at the forward shock. We identify this field as that in the
post shock gas of the forward shock front. 

\subsection{Magnetic field amplification (MFA)}
\label{Su:MFA}
Eq.~(\ref{Eq:MFo}) shows that the MF substantially exceeds
typical stellar wind fields at these distances. For comparison the
equipartition MF in the progenitor wind is \cite{fb98}:
\begin{equation}
\label{Eq:Beq}
B_{eq} = {\left(\dot{M} u_{w}\right)^{1/2} \over r} \nonumber 
\simeq [2.5 \ \rm{mG}] \times \dot{M}_{-5}^{1/2} \times u_{w,10}^{1/2} \times r_{16}^{-1} \ ,
\end{equation}
about one thousand times less than the value in Eq.\ref{Eq:MFo}.  T09
find that the sub-shock compression ratio is close to 4. The
compression of the MF given by Eq.~(\ref{Eq:Beq}) cannot therefore
explain the value given by Eq.~(\ref{Eq:MFo}) (see also
\cite{fb98}). {\it We conclude and assume that a strong MF
  amplification process is at work at the forward shock front, driven
  by the diffusive shock acceleration (DSA) of hadrons.}

\subsubsection{Streaming driven instabilities} 

Streaming of cosmic rays ahead of the shock front produces magnetic
fluctuations. The streaming modes can be in resonance (R) with the
energetic particles, i.e. in the high-energy limit they have a
wave-number such that $k \sim \rm{r}_L^{-1}$; or they can be
non-resonant (NR) with a much larger wave-number \cite{bl01,
  ab09}. The NR modes grow the fastest \cite{plm06}. For 1993J the NR
modes grow and produce magnetic fluctuations over intra-day
timescales; we obtain a growth time:
\begin{equation}
\label{Eq:TNR}
\tau_{NR-st} = [0.16~\rm{day}] \times \left({\phi/15 \over (\xi_{CR}/0.05) u_{sh, 93J}^3 \sqrt{n_{93J}}}\right) \nonumber \\
\times E_{PeV} \ t_{days}^{1.17} \ .
\end{equation}
The CR distribution is assumed to scale as $p^{-4}$ over more than 6
orders of magnitude producing $\phi=\ln(p_{max}/p_{inj})=15$, and
$\xi_{CR}=0.05$, so that 5\% of the fluid kinetic energy is imparted
to energetic particles. $p_{max}$ ($p_{inj}$) is the maximum
(injected) particle momentum. The growth timescale has to be shorter
than the advection timescale towards the shock $\tau_{adv} =
\kappa/V_{sh}^2$. The latter is calculated for a diffusion coefficient $\kappa =
\eta \kappa_{B} > \kappa_B$ taken in the background MF (Eq
\ref{Eq:Beq}) and $\kappa_B=c r_L/3$. We find: $\tau_{adv} =
    [0.24~\rm{day}] \times \eta \ E_{PeV} \times t_{day}^{1.17} $ The
    condition $\tau_{NR-st} < \tau_{adv}$ is necessary but not
    sufficient for the instability to develop. The magnetic
    fluctuations produced by the instability uncovered by
    \cite{bell04} produce small scale perturbations. The wave number
    corresponding to the maximum growth rate is \cite{ab09} $k_{Gmax}
    r_{Lmax} \simeq 4 \times 10^6$ for SN 1993J. Acceleration and
    confinement of energetic particles up to a few PeV requires
    magnetic fluctuations to be generated at resonant scales $k r_L
    \simeq 1$, as the NR growth rate scales as $k^{1/2}$ the MF at the
    scale of interest grows over a timescales about 2000 times larger
    than the one obtained in Eq. \ref{Eq:TNR}. If the background MF is
    purely toroidal $\eta < 1$, and the instability may not have time
    to develop.

The R instability can build up over the NR one \cite{plm06}, with the
ratio of the magnetic energies reaching $\sqrt{\xi_{CR} c/V_{sh}} \sim
0.95$ in our case. The growth timescale for the R instability is
however longer than the rate given by Eq.~\ref{Eq:TNR}. We have
\cite{ab09} $\tau_{R-st} \simeq \sqrt{\pi \sigma/8}/r_{Lmax}$ with
$\sigma\sim 3 \times 10^{16} \rm{cm^2/s^2}$ in the conditions that
prevail for SN 1993J. This leads to a growth rate $\tau_{R-st} \simeq
16 ~ \tau_{NR-st}$ at 1 PeV.  The NR instability can be driven to
non-linear stages and produce large wave numbers. A typical timescale
of non-linear saturation of the MF is about $5 \times \tau_{NR,st}$
\cite{bsrg13}. Recently \cite{boe11} proposed a ponderomotive
instability that builds up on the magnetic fluctuations by the NR
streaming instability. We can evaluate the growth time-scale of such
long-wavelength modes: $\tau_{LW} = [0.29~\rm{day}] \times
\sqrt{\left({\phi/15 \over \xi_{CR}/0.05}\right)} \nonumber \times {1
  \over \sqrt{u_{sh, 93J}^3 A_{10}}} \times E_{PeV}\; t_{days}$. The
parameter $A= B_{NR}^2/B_0^2 > 1$ is the level of magnetic energy
produced by the small scale instability with respect to the background
CSM and is normalized to 10. Long wavelengths are produced on
timescales shorter than $\tau_{adv}$ for $k r_{L,max} \simeq 1$.

\subsubsection{Turbulence driven instabilities}
Stellar winds of massive stars are subject to strong fluid
instabilities that can lead to turbulent motions
\cite{vvd08}. Turbulent density and magnetic fluctuations at SN shocks
can result in MF amplification \cite{bjl09, gj07}. In the latter work
the presence of CRs is not necessary to produce the magnetic field
amplification. The MF growth time is controlled by the coherence
length of the turbulent spectrum L and the fluid velocity
$u_{sh}$. For SN1993J, $L/u_{sh} \sim [0.3~\rm{year}] \ L_{0.01
  pc}/u_{sh,93J}$ hence the magnetic field has to grow over a fraction
of $10^{-2}$ of this timescale.

\subsection{Maximum cosmic ray energies}
\label{S:Max}
Immediately after SN outburst, the maximum energy is likely limited by
the SNR age. Balancing the age with the acceleration time gives
$\tau_{acc} = g(r) \kappa_u/u_{sh}^2$, with $r$ the shock compression
ratio, and $g(r)=3r/(r-1) \times (\kappa_d/\kappa_u r +1)$.  We use
$\kappa_d=\kappa_u/\sqrt{11}$, corresponding to a tangled MF whose
tangential component is compressed by a factor 4 \cite{mc10}.  The
maximum energy is $E_{max,age,PeV} \simeq {12.3 \over \eta g(r)}
\times (1-t_{day}^{-0.17})$.  But rapidly (see \S \ref{Su:MFA}) the
streaming instability amplifies the MF, and the non-linear process
produces a MF at saturation. The typical saturation value
\cite{bell04} is:
\begin{equation}
\label{Eq:Bsat}
B_{sat}= [16~\rm{Gauss}] \times \sqrt{\xi_{CR}/0.05 \over \phi/15} \times t_{days}^{-1} \ .
\end{equation}
This value is within a factor of 2 of the MF derived in the upstream
medium from Eq.\ref{Eq:MFo} using a compression ratio $r=4$. If only
the NR instability is at work building the MF, the maximum particle
energy is then fixed by a condition over the CR areal charge
\cite{bsrg13} that produces $\int \tau_{NR-st}^{-1} dt = 6.8$. The
latter value corresponds to the amplification of the equipartition MF
to the value deduced from radio observations. In that case:
$E_{max,NR,PeV} \sim 1 \times t_{day}^{-0.17}$.  In the case of long
wavelengths fluctuations produced by the ponderomotive instability,
the maximum energy is fixed by geometrical losses. The maximum
energies are obtained with a diffusion coefficient expressed in the
amplified field and compared to $\eta_{esc} R_{sh} u_{sh}$. In order
to derive a time dependence of the maximum energy the time dependence
of the amplified field has to be specified, for which we rely on the
estimate given in Eq. \ref{Eq:Bsat}. This gives:
$E_{max,LW,PeV} \sim 55 \left({\eta_{esc} \over 0.1}\right) \times t_{day}^{-0.34}$.
Apart from the time dependence of $B_{sat}$, this value is optimistic
since the highest energy particles may feel a MF lower than $B_{sat}$
\cite{pzs10}. The maximum cosmic ray (proton) energy at any given time
is the minimum of all the above limits. In all cases, PeV energies are
possible days after the outburst. 

\subsection{Cosmic-ray spectral evolution}
\label{S:CRs}
We follow the prescription of T09 regarding the time evolution of the
CR energy content. The proton particle spectrum is assumed to follow a
power-law with a spectral index $s = 2$ and an exponential cutoff at
$E_{max}(t)$, the maximum particle energy as discussed in \S
\ref{S:Max}. In the case of secondary electrons and positrons produced
in the proton-proton interactions, we solve a one-zone energy equation
to calculate the time evolution of their energy distribution $N(E,t)$:
$\partial_t N(E,t) + \partial_E (\dot{L}(E) N(E,t)) = Q(E,t)$, where
$\dot{L}$ includes the synchrotron losses for secondary electrons and
positrons in the post-shock region.

\section{Gamma-Ray production}
\label{S:GRa}
We derive multi-wavelength time dependent spectra, in the context of
MFA, driven by the instabilities discussed in section \S
\ref{Su:MFA}. We consider only proton-proton interactions, as inverse
Compton or bremsstrahlung radiation have been found negligible in the
GeV-TeV range explored here. Inverse Compton process is highly
disfavored due to the strong magnetic field at the forward
shock. Gamma-rays can be absorbed by different soft photon fields to
produce electron-positron pairs. The main photon source is the SN
photosphere, described in the case of SN 1993J by \cite{lewisetal94}.

\subsection{Gamma-gamma absorption}
We have performed a full calculation of the gamma-gamma opacity
$\tau_{\gamma\gamma}$ including geometrical effects due to the
anisotropic interaction (Renaud et al. 2015, in prep). The final
gamma-ray flux is the unabsorbed flux $F_{\nu, un}$ times an
attenuation factor
$\exp(-\tau_{\gamma\gamma}(E_{\gamma}))$. Gamma-gamma absorption is
strong just after the outburst as the interaction in nearly isotropic,
but thereafter drops as the ratio of the forward shock radius to the
photosphere radius reaches $\sim 3$, which happens after $\sim 5$
days.

\subsection{Cherenkov Telescope Array detectability}
\label{S:MWs}
The time dependent gamma-ray spectra in the very-high energy (100 GeV
$<$ E $<$ 100 TeV) gamma-ray domain are displayed in Figure 1. A
source like SN 1993J would be easily detected by CTA above 1 TeV in 20
h of observing time. {\bf The best time window to detect a gamma-ray
  signal is between a week and a month after the outburst}. Prior to a
week the source is optically thick to gamma-rays, but the gamma-gamma
opacity decreases rapidly due to anisotropic effects.  After a month
the gamma-ray signal becomes too faint due to the decrease in density
of the medium.

\begin{figure}[htbp]
\includegraphics[width=0.85\textwidth]{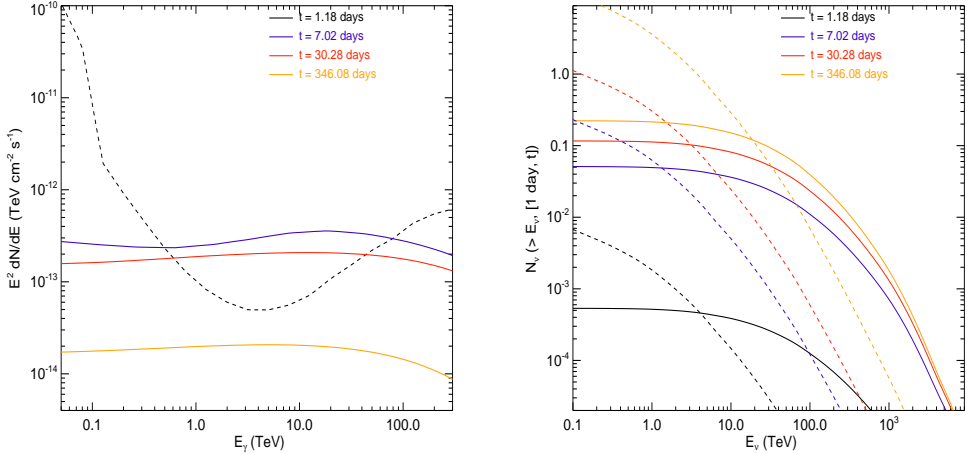}
\caption{[left] Time dependent spectra at 4 different times after
  outburst. Dotted lines - expected CTA sensitivity in 50h. At t=1.18
  days, the source is optically thick to gamma rays, and the flux is
  below the given range. [Right] Time dependent neutrinos flux above
  an energy E expected by a KM3NeT-like instrument (continuous lines)
  at 4 different times after outburst. Dotted lines - atmospheric
  neutrino backgrounds.}
\label{F:Gam}
\end{figure}

\subsection{Neutrino signal}
\label{S:NEu}
Neutrinos are by-products of pion production. The expected flux of
neutrinos detected by an instrument equivalent to KM3NeT is displayed
in Figure 1. We find that at best about 0.1 neutrinos could be
expected from a SN 1993J type event above 10 TeV by summing the
spectra between 1 and 30.28 days after the outburst, with 0.03
background neutrinos. A signal of one neutrino from such a source
requires a source within $\sim$ 1 Mpc, or a gamma-ray signal ten times
stronger.

\section{Discussion \& Conclusions}
\label{S:Dis}
SNe produce multi-PeV particles through the combination of: fast
shocks (v $\sim$ 0.05c), high density CSM produced by stellar winds,
and low wind magnetizations. High degree of CSM ionization eases the
particle acceleration process. Assuming that the background MF has a
turbulent component, different instabilities driven by the
acceleration process can grow over intra-day timescales. This model is
consistent with the strength of the MF seen in SN 1993J.

Parameters affecting the early gamma-ray emission from SNe include the
ratio of the mass loss rate to the wind velocity ($\dot{M}/v_{w}$),
which fixes the CSM medium density and affects the CR driven
instability growth rate. The shock velocity controls the growth rate
of the instabilities and the acceleration timescale. The degree of
ionization \cite{vvd14} is important for the particle acceleration
efficiency and may also produce element dependent CR spectra if the
ionization is partial. The background MF controls partly the local
magnetization and the shock obliquity. The SN luminosity controls the
gamma-gamma absorption process.

Only about 5-6\% of the local core-collapse SNe have been classified
as Type IIb such as SN 1993J \cite{smartt09}. There appear to be two classes of IIb SNe, with compact and extended progenitors \cite{cs10}. It is likely that only the ones with extended radii and higher mass-loss rates such as 1993J are the more likely candidates for detecting $\gamma$-ray emission, thus further reducing the sample. Type IIn SNe are probably more promising targets for gamma-ray telescopes in terms of high ambient density without significantly reduced velocities. In the  case of the IIn SN 1996cr, it has been deduced that the shock was interacting with a shell of density $\sim 10^5\, {\rm cm}^{-3}$ \cite{ddb10} a few years after explosion. This provides a high density target for producing $\gamma$-rays via pion production (although perhaps not as dense as suggested by some authors \cite{muraseetal11} for other SNe).  Such Type IIn SNe would probably also be promising targets for detecting neutrino emission from secondaries. Unfortunately these sources are even less frequent than the 
IIb SNe. SN IIP comprise the largest class of core-collapse SNe, making up around half the total. They
arise from RSG stars, which have wind mass-loss rates ranging from
$10^{-7}$ to $10^{-4}$ M$_{\odot}$ yr$^{-1}$ \cite{mj11}. However, observationally
most IIPs appear to arise from the lower end of RSG stars
\cite{smartt09} and are less luminous in X-rays \cite{vvd14},
suggesting that only the RSG stars with initial masses below $\sim$ 17-19 M$_{\odot}$, with correspondingly lower mass-loss rates \cite{mj11}, explode to become Type IIP SNe. The rare Ib/Ic SNe
harbor the fastest shock waves, but arise from Wolf-Rayet progenitors,
which have wind velocities two orders of magnitude greater than RSGs,
and therefore should have a correspondingly lower wind density. On the other hand, their X-ray flux is presumed to be due to Inverse Compton or synchrotron emission \cite{cf06}, suggesting accelerated electrons, so it is possible that the shocks are capable of also accelerating protons to high energies.  W-R stars are surrounded by low density wind-blown bubbles bordered by a high density shell. If the shell is formed soon before the explosion, as is the case of the SN 2006jc \cite{foley2007}, then it provides a good target for accelerated protons to collide with. Such W-R stars may be good candidates for detecting gamma-ray emission in the early phases. There may be SNe similar to SN 1987A, whose progenitor, a blue-supergiant, had a very low mass-loss rate wind \cite{cd95}, but which shows evidence for a dense HII region with density of order 200 particles cm$^{-3}$ \cite{deweyetal12}, surrounded by a dense circumstellar ring with density $\sim$ 10$^4$ particles cm$^{-3}$ at a distance of $\sim 0.2$ pc from the SN. Finally, the class of super-luminous SNe, especially those that are H-rich \cite{nicholletal15}, may be interacting with extremely dense environments. High densities close in to the star could again provide target material for proton-proton collisions and detectable $\gamma$-ray emission at an early age. \ \
\ \\

{\bf Acknowledgements} This research collaboration is supported by a grant from the FACCTS program to the University of Chicago (PI: VVD; Co-I: MR, Univ of Montpellier). We are grateful to this program for funding travel between Chicago and Montpellier for VVD and MR.

\bibliographystyle{JHEP} 
\bibliography{paper}

\providecommand{\href}[2]{#2}\begingroup\raggedright\begin{thebibliography}{10}

\bibitem{blasi13}
P.~{Blasi}, {\it {The origin of galactic cosmic rays}},  {\em \aapr} {\bf 21}
  (Nov., 2013) 70, [\href{http://arxiv.org/abs/1311.7346}{{\tt
  arXiv:1311.7346}}].

\bibitem{hillas84}
A.~M. {Hillas}, {\it {The Origin of Ultra-High-Energy Cosmic Rays}},  {\em
  \araa} {\bf 22} (1984) 425--444.

\bibitem{bykov01}
A.~M. {Bykov}, {\it {Particle Acceleration and Nonthermal Phenomena in
  Superbubbles}},  {\em \ssr} {\bf 99} (Oct., 2001) 317--326.

\bibitem{parizot04}
E.~{Parizot}, A.~{Marcowith}, E.~{van der Swaluw}, A.~M. {Bykov}, and
  V.~{Tatischeff}, {\it {Superbubbles and energetic particles in the Galaxy. I.
  Collective effects of particle acceleration}},  {\em \aap} {\bf 424} (Sept.,
  2004) 747--760, [\href{http://arxiv.org/abs/astro-ph/0405531}{{\tt
  astro-ph/0405531}}].

\bibitem{ferrand10}
G.~{Ferrand} and A.~{Marcowith}, {\it {On the shape of the spectrum of cosmic
  rays accelerated inside superbubbles}},  {\em \aap} {\bf 510} (Feb., 2010)
  A101, [\href{http://arxiv.org/abs/0911.4457}{{\tt arXiv:0911.4457}}].

\bibitem{vt09}
V.~{Tatischeff}, {\it {Radio emission and nonlinear diffusive shock
  acceleration of cosmic rays in the supernova SN 1993J}},  {\em \aap} {\bf
  499} (May, 2009) 191--213, [\href{http://arxiv.org/abs/0903.2944}{{\tt
  arXiv:0903.2944}}].

\bibitem{ms09}
J.~R. {Maund} and S.~J. {Smartt}, {\it {The Disappearance of the Progenitors of
  Supernovae 1993J and 2003gd}},  {\em Science} {\bf 324} (Apr., 2009) 486--,
  [\href{http://arxiv.org/abs/0903.3772}{{\tt arXiv:0903.3772}}].

\bibitem{foxetal14}
O.~D. {Fox}, K.~{Azalee Bostroem}, S.~D. {Van Dyk}, A.~V. {Filippenko},
  C.~{Fransson}, T.~{Matheson}, S.~B. {Cenko}, P.~{Chandra}, V.~{Dwarkadas},
  W.~{Li}, A.~H. {Parker}, and N.~{Smith}, {\it {Uncovering the Putative B-star
  Binary Companion of the SN1993J Progenitor}},  {\em ApJ} {\bf 790} (July,
  2014) 17, [\href{http://arxiv.org/abs/1405.4863}{{\tt arXiv:1405.4863}}].

\bibitem{flc96}
C.~{Fransson}, P.~{Lundqvist}, and R.~A. {Chevalier}, {\it {Circumstellar
  Interaction in SN 1993J}},  {\em \apj} {\bf 461} (Apr., 1996) 993--+.

\bibitem{chandraetal04}
P.~{Chandra}, A.~{Ray}, and S.~{Bhatnagar}, {\it {The Late-Time Radio Emission
  from SN 1993J at Meter Wavelengths}},  {\em \apj} {\bf 612} (Sept., 2004)
  974--987, [\href{http://arxiv.org/abs/astro-ph/0405448}{{\tt
  astro-ph/0405448}}].

\bibitem{fb98}
C.~{Fransson} and C.~{Bj{\"o}rnsson}, {\it {Radio Emission and Particle
  Acceleration in SN 1993J}},  {\em \apj} {\bf 509} (Dec., 1998) 861--878,
  [\href{http://arxiv.org/abs/astro-ph/}{{\tt astro-ph/}}].

\bibitem{bl01}
A.~R. {Bell} and S.~G. {Lucek}, {\it {Cosmic ray acceleration to very high
  energy through the non-linear amplification by cosmic rays of the seed
  magnetic field}},  {\em MNRAS} {\bf 321} (Mar., 2001) 433--438.

\bibitem{ab09}
E.~{Amato} and P.~{Blasi}, {\it {A kinetic approach to cosmic-ray-induced
  streaming instability at supernova shocks}},  {\em \mnras} {\bf 392} (Feb.,
  2009) 1591--1600, [\href{http://arxiv.org/abs/0806.1223}{{\tt
  arXiv:0806.1223}}].

\bibitem{plm06}
G.~{Pelletier}, M.~{Lemoine}, and A.~{Marcowith}, {\it {Turbulence and particle
  acceleration in collisionless supernovae remnant shocks. I. Anisotropic
  spectra solutions}},  {\em \aap} {\bf 453} (July, 2006) 181--191,
  [\href{http://arxiv.org/abs/astro-ph/0603461}{{\tt astro-ph/0603461}}].

\bibitem{bell04}
A.~R. {Bell}, {\it {Turbulent amplification of magnetic field and diffusive
  shock acceleration of cosmic rays}},  {\em \mnras} {\bf 353} (Sept., 2004)
  550--558.

\bibitem{bsrg13}
A.~R. {Bell}, K.~M. {Schure}, B.~{Reville}, and G.~{Giacinti}, {\it {Cosmic-ray
  acceleration and escape from supernova remnants}},  {\em \mnras} {\bf 431}
  (May, 2013) 415--429, [\href{http://arxiv.org/abs/1301.7264}{{\tt
  arXiv:1301.7264}}].

\bibitem{boe11}
A.~M. {Bykov}, S.~M. {Osipov}, and D.~C. {Ellison}, {\it {Cosmic ray current
  driven turbulence in shocks with efficient particle acceleration: the
  oblique, long-wavelength mode instability}},  {\em \mnras} {\bf 410} (Jan.,
  2011) 39--52, [\href{http://arxiv.org/abs/1010.0408}{{\tt arXiv:1010.0408}}].

\bibitem{vvd08}
V.~V. {Dwarkadas}, {\it {Turbulence in wind-blown bubbles around massive
  stars}},  {\em Physica Scripta Volume T} {\bf 132} (Dec., 2008) 014024,
  [\href{http://arxiv.org/abs/0810.4361}{{\tt arXiv:0810.4361}}].

\bibitem{bjl09}
A.~{Beresnyak}, T.~W. {Jones}, and A.~{Lazarian}, {\it {Turbulence-Induced
  Magnetic Fields and Structure of Cosmic Ray Modified Shocks}},  {\em \apj}
  {\bf 707} (Dec., 2009) 1541--1549,
  [\href{http://arxiv.org/abs/0908.2806}{{\tt arXiv:0908.2806}}].

\bibitem{gj07}
J.~{Giacalone} and J.~R. {Jokipii}, {\it {Magnetic Field Amplification by
  Shocks in Turbulent Fluids}},  {\em \apjl} {\bf 663} (July, 2007) L41--L44.

\bibitem{mc10}
A.~{Marcowith} and F.~{Casse}, {\it {Postshock turbulence and diffusive shock
  acceleration in young supernova remnants}},  {\em \aap} {\bf 515} (June,
  2010) A90, [\href{http://arxiv.org/abs/1001.2111}{{\tt arXiv:1001.2111}}].

\bibitem{pzs10}
V.~{Ptuskin}, V.~{Zirakashvili}, and E.-S. {Seo}, {\it {Spectrum of Galactic
  Cosmic Rays Accelerated in Supernova Remnants}},  {\em \apj} {\bf 718} (July,
  2010) 31--36, [\href{http://arxiv.org/abs/1006.0034}{{\tt arXiv:1006.0034}}].

\bibitem{lewisetal94}
J.~R. {Lewis}, N.~A. {Walton}, W.~P.~S. {Meikle}, R.~{Martin}, R.~J. {Cumming},
  R.~M. {Catchpole}, M.~{Arevalo}, R.~W. {Argyle}, C.~R. {Benn}, P.~S.
  {Bunclark}, and H.~O. e.~a. {Castaneda}, {\it {Optical Observations of
  Supernova 1993J from La-Palma - Part One - Days 2 TO 125}},  {\em \mnras}
  {\bf 266} (Jan., 1994) L27.

\bibitem{vvd14}
V.~V. {Dwarkadas}, {\it {On the lack of X-ray bright Type IIP supernovae}},
  {\em \mnras} {\bf 440} (May, 2014) 1917--1924,
  [\href{http://arxiv.org/abs/1402.5150}{{\tt arXiv:1402.5150}}].

\bibitem{smartt09}
S.~J. {Smartt}, {\it {Progenitors of Core-Collapse Supernovae}},  {\em \araa}
  {\bf 47} (Sept., 2009) 63--106, [\href{http://arxiv.org/abs/0908.0700}{{\tt
  arXiv:0908.0700}}].

\bibitem{cs10}
R.~A. {Chevalier} and A.~M. {Soderberg}, {\it {Type IIb Supernovae with Compact
  and Extended Progenitors}},  {\em \apjl} {\bf 711} (Mar., 2010) L40--L43,
  [\href{http://arxiv.org/abs/0911.3408}{{\tt arXiv:0911.3408}}].

\bibitem{ddb10}
V.~V. {Dwarkadas}, D.~{Dewey}, and F.~{Bauer}, {\it {Bursting SN 1996cr's
  bubble: hydrodynamic and X-ray modelling of its circumstellar medium}},  {\em
  \mnras} {\bf 407} (Sept., 2010) 812--829,
  [\href{http://arxiv.org/abs/1005.1090}{{\tt arXiv:1005.1090}}].

\bibitem{muraseetal11}
K.~{Murase}, T.~A. {Thompson}, B.~C. {Lacki}, and J.~F. {Beacom}, {\it {New
  class of high-energy transients from crashes of supernova ejecta with massive
  circumstellar material shells}},  {\em PRD} {\bf 84} (Aug., 2011) 043003,
  [\href{http://arxiv.org/abs/1012.2834}{{\tt arXiv:1012.2834}}].

\bibitem{mj11}
N.~{Mauron} and E.~{Josselin}, {\it {The mass-loss rates of red supergiants and
  the de Jager prescription}},  {\em \aap} {\bf 526} (Feb., 2011) A156,
  [\href{http://arxiv.org/abs/1010.5369}{{\tt arXiv:1010.5369}}].

\bibitem{cf06}
R.~A. {Chevalier} and C.~{Fransson}, {\it {Circumstellar Emission from Type Ib
  and Ic Supernovae}},  {\em \apj} {\bf 651} (Nov., 2006) 381--391,
  [\href{http://arxiv.org/abs/astro-ph/}{{\tt astro-ph/}}].

\bibitem{foley2007}
R.~J. {Foley}, N.~{Smith}, M.~{Ganeshalingam}, W.~{Li}, R.~{Chornock}, and
  A.~V. {Filippenko}, {\it {SN 2006jc: A Wolf-Rayet Star Exploding in a Dense
  He-rich Circumstellar Medium}},  {\em \apjl} {\bf 657} (Mar., 2007)
  L105--L108.

\bibitem{cd95}
R.~A. {Chevalier} and V.~V. {Dwarkadas}, {\it {The Presupernova H II Region
  around SN 1987A}},  {\em \apjl} {\bf 452} (Oct., 1995) L45--+.

\bibitem{deweyetal12}
D.~{Dewey}, V.~V. {Dwarkadas}, F.~{Haberl}, R.~{Sturm}, and C.~R. {Canizares},
  {\it {Evolution and Hydrodynamics of the Very Broad X-Ray Line Emission in SN
  1987A}},  {\em \apj} {\bf 752} (June, 2012) 103,
  [\href{http://arxiv.org/abs/1111.5314}{{\tt arXiv:1111.5314}}].

\bibitem{nicholletal15}
M.~{Nicholl}, S.~J. {Smartt}, A.~{Jerkstrand}, and C.~{Inserra}, {\it {On the
  diversity of super-luminous supernovae: ejected mass as the dominant
  factor}},  {\em ArXiv e-prints} (Mar., 2015)
  [\href{http://arxiv.org/abs/1503.0331}{{\tt arXiv:1503.0331}}].

\end{thebibliography}\endgroup


\end{document}